\documentclass[12pt]{article}
\usepackage{amstex}
\begin{document}

\def\vev#1{\langle #1 \rangle}
\def\tr{\mbox{Tr}\,}
\def\ad{\mbox{ad}\,}
\def\ker{\mbox{Ker}\,}
\def\im{\mbox{Im}\,}
\def\kbar{{\mathchar'26\mkern-9muk}}  

\title{Fuzzy Surfaces of Genus Zero}

\author{J. Madore \\
        Laboratoire de Physique Th\'eorique et Hautes 
        Energies\thanks{Laboratoire associ\'e au CNRS, URA D0063}\\
        Universit\'e de Paris-Sud, B\^at. 211, F-91405 Orsay\\       
       }

\date{July, 1997}

\maketitle

\abstract{A fuzzy version of the ordinary round 2-sphere  has been 
constructed with an invariant curvature.  We here consider linear
connections on arbitrary fuzzy surfaces of genus zero.  We shall find as
before that they are more or less rigidly dependent on the differential
calculus used but that a large number of the latter can be constructed
which are not covariant under the action of the rotation group. For
technical reasons we have been forced to limit our considerations to
fuzzy surfaces which are small perturbations of the fuzzy sphere.}

\vfill
\noindent
LPTHE Orsay 97/26
\bigskip
\eject

\parskip 4pt plus2pt minus2pt

\section{Introduction}

A fuzzy version of the ordinary round 2-dimensional sphere $S^2$ has
been given (Madore 1992, Madore \& Grosse 1992, Grosse \& Pre\v snajder
1993, Grosse, Klim\v c\'\i k \& Pre\v snajder 1997b, Carow-Watamura \&
Watamura 1997) which uses a noncommutative geometry based on a
filtration of the algebra $M_n$ of $n \times n$ complex matrices and a
differential calculus based on a set of derivations which form the
irreducible representation of dimension $n$ of the Lie algebra of
$SU_2$. The rotation group acts on this fuzzy sphere and the only
torsion-free metric connection on it has an invariant `Gaussian'
curvature.  Recently Dimakis \& Madore (1996) have proposed a method of
constructing an arbitrary number of differential calculi over a given
noncommutative associative algebra. We shall use this method to
construct linear connections on an arbitrary fuzzy surface $\Sigma$ of
genus zero. We shall find as before that the connections are more or
less rigidly dependent on the differential calculus but that a large
number of the latter can be constructed associated to metrics which are
not covaraiant under the action of the rotation group. For technical
reasons we have been forced to limit our considerations to fuzzy
surfaces which are small perturbations of the fuzzy sphere.

There seems to be an intimate connection between the commutation
relations of an algebra and the linear curvature which it can support.
Evidence of this is to be found in previous articles (Madore 1992,
Dimakis \& Madore 1996, Madore \& Mourad 1996, Madore 1996, 1997).  To a
certain extent the present results furnish another example.

In Section~2 we recall the general method for constructing differential
calculi based on a set of derivations which do not necessarily close to
form a Lie algebra and we briefly recall the definition of a linear
connection which we shall use. Our calculations are based on the
noncommutative equivalent of the moving-frame formalism. In Section~3
we construct the frames and in Section~4 we calculate the associated
torsion-free metric connections.

\section{General Formalism}

Suppose one is given an $M_n$-bimodule $\Omega^1(M_n)$ and an application
$$
M_n \buildrel d \over \longrightarrow \Omega^1(M_n)                 \eqno(2.1)
$$
of $M_n$ into $\Omega^1(M_n)$. Then there is a construction
(Connes 1994, Dimakis \& Madore 1996, Madore \& Mourad 1996) which
yields a differential calculus $\Omega^*(M_n)$ over $M_n$ which has
$\Omega^1(M_n)$ as module of 1-forms. We construct (2.1) using a set of
derivations (Dubois-Violette 1988, Dubois-Violette {\it et al.} 1989,
Dimakis \& Madore 1996). Let $\lambda_a$, $1 \leq a \leq 3$, be a set of
linearly independent anti-hermitian elements of $M_n$ such
that only matrices proportional to the identity commutate with it.
The $*$-algebra generated by the $\lambda_a$ is equal
then to $M_n$.  Consider the derivations $e_a = \ad \lambda_a$. In order
for them to have the correct dimensions one must introduce a mass
parameter $\mu$ and replace $\lambda_a$ by $\mu \lambda_a$. We shall set
$\mu = 1$.  Define the map (2.1) by
$$
df (e_a) = e_a \, f.                                                \eqno(2.2)
$$

Under certain conditions, given below, $\Omega^1(M_n)$ has a basis
$\theta^a$ dual to the derivations,
$$
\theta^a(e_b) = \delta^a_b,                                         \eqno(2.3)
$$
a basis which commutes therefore with the elements $f$ of $M_n$:
$$
f \theta^a = \theta^a f.                                            \eqno(2.4)
$$
In such cases $\Omega^1(M_n)$ is free as a left or right module and can
be identified with the direct sum of 3 copies of $M_n$:
$$
\Omega^1(M_n) = M_n \oplus M_n \oplus M_n.                          \eqno(2.5)
$$

The $\theta^a \otimes \theta^b$ form a basis for 
$\Omega^1(M_n) \otimes_{M_n} \Omega^1(M_n)$. The module $\Omega^2(M_n)$
of 2-forms can be identified as a submodule of 
$\Omega^1(M_n) \otimes_{M_n} \Omega^1(M_n)$ and the projection onto the
2-forms can be written in the form
$$
\pi(\theta^a \otimes \theta^b) = 
P^{ab}{}_{cd} \theta^c \otimes \theta^d                             \eqno(2.6)
$$
where the $P^{ab}{}_{cd}$ are complex numbers which satisfy
$$
P^{ab}{}_{ef} P^{ef}{}_{cd} =  P^{ab}{}_{cd}.                       \eqno(2.7)
$$
The product $\theta^a \theta^b$ satisfies therefore
$$
\theta^a \theta^b = P^{ab}{}_{cd} \theta^c \theta^d.                \eqno(2.8)
$$
If the $\theta^a$ exist then it can be shown (Dimakis \& Madore 1996, 
Madore \& Mourad 1996) that the $\lambda_a$ must satisfy the equation
$$
2 \lambda_c \lambda_d P^{cd}{}_{ab} - 
\lambda_c F^c{}_{ab} - K_{ab} = 0                                   \eqno(2.9)
$$
with $F^c{}_{ab}$ and $K_{ab}$ complex numbers. 

The structure elements $C^a{}_{bc}$ are defined by the equation
$$
d\theta^a = 
- {1\over 2} C^a{}_{bc} \theta^b \theta^c.                         \eqno(2.10)
$$
They are related to the coefficients of (2.9) by
$$
C^a{}_{bc} = F^a{}_{bc} - 2 \lambda_d P^{(da)}{}_{bc}.             \eqno(2.11)
$$

The fuzzy sphere has a differential calculus defined by choosing the
$\lambda_a$ to be the irreducible $n$-dimensional representation of the
Lie algebra of $SU_2$. In this case
$$
P^{ab}{}_{cd} = 
{1\over 2} (\delta^a_c \delta^b_d - \delta^b_c \delta^a_d),        \eqno(2.12)
$$
the $C^a{}_{bc} = F^a{}_{bc}$ are the $SU_2$ structure constants and the
$K_{bc}$ vanish. 

To discuss the commutative limit it is convenient to change the
normalization of the generators $\lambda_a$. We introduce the parameter
$\kbar$ with the dimensions of (length)$^2$ and define `coordinates'
$x_a$ by
$$
x_a = i \kbar \lambda_a.                                           \eqno(2.13)
$$
The $x_a$ satisfy therefore the commutation relations
$$
[x_a , x_b ] = i \kbar x_c C^c{}_{ab}.                             \eqno(2.14)
$$
We choose the $\lambda_a$ so that $C_{abc} = r^{-1} \epsilon_{abc}$
where $r$ is a length parameter. These structure constants are in
general independent from the structure elements introduced in (2.10) but
in the case of the fuzzy sphere they are equal.  Introduce the
$SU_2$-Casimir metric $g_{ab}$. The matrix $g^{ab} x_a x_b$ is the Casimir
operator. We choose $\kbar$ so that $g^{ab} x_a x_b = r^2$. We have then
from (2.14)
$$
4 r^4 = (n^2-1) \kbar^2.                                           \eqno(2.15)
$$
The commutative limit is the limit $\kbar \rightarrow 0$. Were we
considering a noncommutative model of space-time then we would be
tempted to identify $\kbar$ with the inverse of the square of the Planck
mass, $\kbar = \mu^{-2}_P$, and consider space-time as fundamentally
noncommutative in the presence of gravity.

The differential calculus on the fuzzy sphere has (Madore 1992) a basis
$$
\theta^a = - C^a{}_{bc} x^b dx^c - i\kbar r^{-2} \theta x^a.       \eqno(2.16)
$$
The 1-form $\theta$ can be written
$$
\theta = i \kbar^{-1} x_a \theta^a = r^2 \kbar^{-2} x_a dx^a.      \eqno(2.17)
$$
In the commutative limit $\theta$ diverges but 
$\kbar \theta \rightarrow r^2 A$ where $A$ is the Dirac-monopole
potential of unit magnetic charge.  The commutative limit of the frame
$\theta^a$ is a moving frame on a $U_1$-bundle over $S^2$. In this case
it is not the frame bundle. A standard Kaluza-Klein reduction gives rise
to the potential $A$ as well as the geometry of the sphere.

The formalism which we shall use resembles that which can be used to
describe a manifold $V$ of dimension $d$ which is defined by its
metric embedding in a flat euclidean space ${\mathbb R}^n$. If there
are no topological complications then generically $n = d(d+1)/2$ since
this is the number of independent components of a local metric on $V$.
Let $x^a$ be the coordinates of the embedding space and $y^\alpha$ local
coordinates of $V$. Then $V$ is defined locally by equations of the form
$x^a = x^a(y^\alpha)$. Let $g_{ab}$ be the components of the flat metric
on ${\mathbb R}^n$.  The local components $h_{\alpha\beta}$ of the
induced metric on $V$ are given by
$$
h_{\alpha\beta} = g_{ab} {\partial x^a \over \partial y^\alpha}
{\partial x^b \over \partial y^\beta}
$$
Let $x^{\prime a} = x^a + h^a(x^b)$ be a variation of the coordinates of
the embedding space ${\mathbb R}^n$. The most general variation
$h^\prime_{\alpha\beta}$ of $h_{\alpha\beta}$ can be obtained by the
induced variation $g^\prime_{ab} = g_{ab} - \partial_{(a} h_{b)}$ of
$g_{ab}$:
$$
h^\prime_{\alpha\beta} = h_{\alpha\beta} -
\partial_{(a} h_{b)} {\partial x^a \over \partial y^\alpha}
{\partial x^b \over \partial y^\beta}.                             \eqno(2.18)
$$
If $\theta^a$ is a moving frame on ${\mathbb R}^n$ then the new metric
$g^\prime$ is given by
$$
g^\prime (\theta^a \otimes \theta^b) = g^{ab} + \partial^{(a} h^{b)}.
$$
A moving frame for $g^\prime$ is given then by
$\theta^{\prime a} = (\delta^a_b + \tilde \Lambda^a_b) \theta^b$ with
$$
\tilde \Lambda^a_b = - \partial_b h^a.                             \eqno(2.19)
$$

The algebra ${\cal C}(V)$ of smooth functions on $V$ can be identified
with a quotient of the algebra of ${\cal C}({\mathbb R}^n)$ and the
differential calculus $\Omega^*(V)$ can be identified with a quotient of
the calculus $\Omega^*({\mathbb R}^n)$. It is convenient to designate by
$dx^\alpha$ the element $dx^a$ of $\Omega^1({\mathbb R}^n)$ restricted
to $V$. Then $1 \leq \alpha \leq n$, but the differentials $dx^\alpha$
satisfies $n-d$ relations. The projection of $\Omega^1({\mathbb R}^n)$
onto $\Omega^1(V)$ can be written in the form
$$
dx^\alpha = \pi^\alpha_a dx^a.                                     \eqno(2.20)
$$
The moving frame $\theta^a$ can be chosen such that 
$\pi^\alpha_a \theta^a$ is a moving frame on $V$ for the induced metric
with components $h_{\alpha\beta}$. The transformation to a moving frame
for the perturbed metric $h^\prime_{\alpha\beta}$ is given therefore by
$$
\pi^\alpha_a (\delta^a_b + \tilde \Lambda^a_b) = 
\pi^\alpha_b - \pi^\alpha_a \partial_b h^a.                        \eqno(2.21)
$$

There is also a projection of the parallelizable Hopf bundle 
${\cal P}$ onto $\Sigma$.  Therefore ${\cal C}(\Sigma)$ can be
identified as a subalgebra of ${\cal C}({\cal P})$ and the
differential calculus $\Omega^*(\Sigma)$ can be identified with a
differential subalgebra of the calculus $\Omega^*({\cal P})$.
We shall use this identification also when considering the differential
calculus over $\Sigma$. In general there is always a projection of the
(parallelizable) frame bundle ${\cal L}$ of a manifold $V$ onto $V$
itself and so $\Omega^*(V)$ can be considered as a subalgebra of a
differential calculus $\Omega^*({\cal L})$ which has a free module of
1-forms.

\section{Perturbations of the calculus}

We are interested in perturbations of the fuzzy sphere which could be
considered as noncommutative versions of a topological sphere which is
not invariant under the action of the rotation group. We first introduce
a perturbation $\Omega^{\prime *}(M_n)$ of the differential calculus
$\Omega^*(M_n)$ of the fuzzy sphere. We shall require that the perturbed
differential calculus be based on derivations. We introduce then 
$$
e^\prime_a = {1\over i\kbar} \ad x^\prime_a
$$
with
$$
x^\prime_a = x_a + h_a                                             \eqno(3.1)
$$
and where $h_a$ is a set of 3 elements of $M_n$ with 
$\vert h_a \vert << r$.  If we write the perturbed version of (2.9) as
$$
2 x^\prime_c x^\prime_d P^{\prime cd}{}_{ab} - 
i \kbar x^\prime_c F^{\prime c}{}_{ab} + \kbar^2 K^\prime_{ab} = 0
$$
and expand the coefficients as
$$
P^{\prime cd}{}_{ab} = 
{1\over 2} (\delta^c_a \delta^d_b - \delta^d_a \delta^c_b) +
P^{cd}_{(1)ab}, \qquad 
F^{\prime c}{}_{ab} = C^c{}_{ab} + F^c_{(1)ab}, \qquad 
K^\prime_{ab} = K_{(1)ab}
$$
we find the equation
$$
[x_a, h_b] - [x_b, h_a] + 2 x_c x_d P^{cd}_{(1)ab} -
i \kbar h_c C^c{}_{ab} - i \kbar x_c F^c_{(1)ab} + 
\kbar^2 K_{(1)ab} = 0                                              \eqno(3.2)
$$
for the perturbations.  We can suppose that $P^{cd}_{(1)ab}$ is
symmetric in the first two indices since an antisymmetric part can be
absorbed in a redefinition of the fifth term.  It must be then
antisymmetric in the last two if $P^{\prime cd}{}_{ab}$ is to be a 
projector.  The $P^{cd}_{(1)ab}$ could be also chosen to be trace-free
in the first two indices since the trace part could be absorbed in a
redefinition of the last term.

The most general solution to the equation
$$
[x_a, h_b] - [x_b, h_a] - i \kbar h_c C^c{}_{ab} = 0               \eqno(3.3)
$$
is given by $h_a = [x_a, f]$ where $f$ is an arbitrary anti-hermitian
matrix. We obtain thereby $n^2-1$ linearly independent solutions to
Equation~(3.2) with
$$
P^{cd}_{(1)ab} = 0, \qquad F^c_{(1)ab} = 0, \qquad K_{(1)ab} = 0.   \eqno(3.4)
$$
These solutions correspond to the trivial redefinition of the generators
$x_a$ which leave invariant the commutation relations (2.14). The most
general such operation is given by the map $x_a \mapsto u^{-1} x_a u$
with $u$ a unitary matrix. In the linear approximation $u \simeq 1 + f$.
The commutative limit suggests the Ansatz $h_a = h x_a$ with $h$ an
arbitrary element of $M_n$.  Equation~(3.2) simplifies to the equation
$$
x_{[a} h x_{b]} + 2 x_c x_d P^{cd}_{(1)ab} - 
i \kbar x_c F^c_{(1)ab} + \kbar^2 K_{(1)ab} = 0                     \eqno(3.5)
$$
with $P^{cd}_{(1)ab}$ symmetric in the first two indices. We have been
unable to find interesting explicit solutions to this equation.

Some solutions to (3.2) for $h_a$ can be found by inspection.  A trivial
`monopole' solution, with the $h_a = f_a$ all proportional to the unit
matrix, is given by
$$
P^{cd}_{(1)ab} = 0, \qquad F^c_{(1)ab} = 0, \qquad
\kbar K_{(1)ab} = i f_c C^c{}_{ab}.                                 \eqno(3.6)
$$
A `dipole' solution of the form $h_a = f_{ab} x^b$ is given by
$$
P^{cd}_{(1)ab} = 0, \qquad 
F_{(1)abc} = C^d{}_{a[b} f_{c]d} - C^d{}_{bc} f_{da}, \qquad       
K_{(1)ab} = 0.                                                      \eqno(3.7)
$$
A `quadrupole' solution of the form $h_a = f_{abc} x^b x^c$ can also 
be found with $f_{abc}$ real and symmetric and trace-free in the last 
two indices. It is given by
$$
P^{cd}_{(1)ab} = {1\over 2} i \kbar f_{[a}{}^{e(c} C^{d)}{}_{b]e},
\qquad F^c_{(1)ab} = 0, \qquad K_{(1)ab} = 0.                      \eqno(3.8)
$$
Since $C_{abc}$ is completely antisymmetric $P^{cd}_{(1)ab}$ is
trace-free in the first two indices.  From the perturbed version of
Equation~(2.8) we see that since $P^{\prime cd}{}_{ab}$ is not
antisymmetric in the first two indices the $\theta^{\prime a}$ do not
anticommute. From the perturbed version of (2.11) we see that the
structure elements $C^{\prime a}{}_{bc}$ are given by
$$
C^{\prime c}{}_{ab} = C^c{}_{ab} + 
4 i \kbar^{-1} x_ d P^{cd}_{(1)ab}.                               \eqno(3.9)
$$

We have defined the structure of the algebra in terms of solutions to the
Equation~(2.9). It might be also interesting to define it in terms of
the commutation relations for the $x^\prime_a$:
$$
[x^\prime_a, x^\prime_b] = i \kbar x^\prime_c F^{\prime c}{}_{ab} 
- 2 x^\prime_c x^\prime_d (P^{\prime cd}{}_{ab} - P^{cd}{}_{ab})
- \kbar^2 K^\prime_{ab} = 0.
$$
In the linear approximation this becomes for the solution (3.8)
$$
[x^\prime_a, x^\prime_b] = i\kbar x^\prime_c C^c{}_{ab} -
i \kbar x^\prime_c x^\prime_d f_{[a}{}^{e(c} C^{d)}{}_{b]e}.      \eqno(3.10)
$$
The $x^\prime_a$ are a new set of generators of the matrix algebra which
satisfy different commutation relations. The differential calculus is
defined in terms of the corresponding derivations $e^\prime_a$. In the
commutative limit the commutation relations (3.10) induce a Poisson
structure
$$
\{x^\prime_a, x^\prime_b\} = x^\prime_c C^c{}_{ab} -
x^\prime_c x^\prime_d f_{[a}{}^{e(c} C^{d)}{}_{b]e}.              \eqno(3.11)
$$
on the surface. The $x^\prime_a$ designate here coordinates in 
${\mathbb R}^3$. It is interesting to note that although the commutation
relations (3.10) do not in principle fix the noncommutative differential
calculus, the limiting Poisson structure (3.11) depends on the
differential structure of $\Sigma$.  The coordinate transformation 
$x_a \mapsto x^\prime_a$ is not a symplectomorphism. It was shown some
time ago that symplectomorphisms are limits of transformations of the
form (3.3); this was one of the results of the authors quoted below in
Section~5.  We refer to Madore (1995) for a recent discussion.

The condition (2.9) is a necessary condition for the existence of the
dual basis $\theta^a$. To show that effectively the $\theta^{\prime a}$
exist we must construct them explicitly. We write
$$
\theta^{\prime a} = \theta^a + \theta^a_{(1)}.
$$
Then from the perturbed version of (2.3) we find the relation
$$
\theta^a_{(1)} (e_b) + \theta^a (e_{(1)b}) = 0                     \eqno(3.12)
$$
for the perturbation $\theta^a_{(1)}$. We write this as
$$
i \kbar \theta^a_{(1)} (e_b) = C^a{}_{cd} x^c [h_b, x^d] + 
i \kbar^{-1} x_c [h_b, x^c] x^a.                                   \eqno(3.13)
$$
We shall find below that the $f_{abc}$ must be completely symmetric in
all indices. In this case we find that
$$
\theta^a_{(1)} = - (2 - {\kbar^2\over r^4}) f^a{}_{bc} x^b \theta^c     
- r^{-2} h_b x^a \theta^b +
i \kbar r^{-2} C^a{}_{bc} f^b{}_{de} x^d x^c \theta^e.             \eqno(3.14)
$$
We can write therefore 
$$
\theta^{\prime a} = (\delta^a_b + \Lambda^a_b) \theta^b            \eqno(3.15)
$$
where 
$$
\Lambda^a_b = - (2 - {\kbar^2\over r^4}) f^a{}_{bc} x^c     
- r^{-2} h_b x^a  + i \kbar r^{-2} C^a{}_{cd} f^c{}_{eb} x^e x^d.  \eqno(3.16)
$$

It must be emphasized that there are two distinct differential calculi
and two distinct modules of 1-forms. In $\Omega^1(M_n)$ the $\theta^a$
are the preferred basis which commute with the elements of the algebra.
The $\theta^{\prime a}$ satisfy therefore from (3.15) the commutation
relations 
$$
[\theta^{\prime a}, f] = [\Lambda^a_b, f] \,\theta^b \neq 0.
$$
In $\Omega^{\prime 1}(M_n)$ the $\theta^{\prime a}$ are the preferred
basis which commute with the elements of the algebra.  The $\theta^a$
satisfy therefore from (3.15) the commutation relations
$$
[\theta^a, f] = - [\Lambda^a_b, f] \,\theta^{\prime b} \neq 0.
$$
Since the $\theta^a$ are defined on the derivations $e_a$ and the
$\theta^{\prime a}$ are defined on the derivations $e^\prime_a$, to
compare the two we must extend them both to the union of the two sets of
derivations.  This can be done since the differential $dx^a$ can be
defined on an arbitrary derivation $X$ by the formula $dx^a(X) = X x^a$.
On the extended set neither basis commutes with the elements of the 
algebra. We recall that the vector space of all derivations of $M_n$ is of
dimension $n^2-1$ and the modules of 1-forms we consider here are defined
on subspaces thereof of dimension 3.

In the commutative limit both the differential calculi defined above
tend to the de~Rham differential calculus on the $U_1$-bundle ${\cal P}$
over $\Sigma$. In this limit $\Lambda^a_b \rightarrow \tilde \Lambda^a_b$
where
$$
\tilde \Lambda^a_b = - 2 f^a{}_{bc} x^c - r^{-2} h_b x^a =
- \partial_b h^a - r^{-2} h_b x^a                                 \eqno(3.17)
$$
is a change of basis between two moving frames on ${\cal P}$. The sphere
$S^2$ is defined by the relation $g_{ab} x^a x^b - r^2 = 0$ on the ordinary
euclidian coordinates of ${\mathbb R}^3$. We have therefore (2.20) with
$$
\pi^\alpha_a = \delta^\alpha_a - r^{-2} x^\alpha x_a
$$
and (2.21) follows from (3.17). One verifies that $\pi^\alpha_a \theta^a$
is, by restriction to $\Omega^1 (S^2)$, a (redundent) moving 
frame for the metric on the sphere.  The perturbed metric is given
therefore by (2.21).

The quadrupole perturbation is rather special.  However it can be
combined with a redefinition of the generators given by a
nonperturbative matrix $u$.  In this way an arbitrarily large number of
perturbative solutions to the Equation~(2.9) can be found. It is easy to
localize the obstruction to higher-order multipole perturbations in the
linear approximation. The third term in Equation~(3.2) is only
quadratic in the variables and cannot therefore combine with higher-order
polynomials appearing in the first two terms. Such a polynomial
would appear only in the next approximation with terms of the form 
$x_{(c} h_{d)} P^{cd}_{(1)ab}$.

\section{Linear connections}

A linear connection is defined (Dubois-Violette {\it et al.} 1995, 1996) 
as a couple $(D, \sigma)$ where $D$ is a covariant derivative and
$\sigma$ is a generalized permutation. The unique torsion-free metric
connection on the fuzzy sphere, or rather the fuzzy $U_1$-bundle over
it, is given by
$$
D\theta^a = - {1\over 2} C^a{}_{bc} \theta^b \otimes \theta^c, \qquad
\sigma(\theta^a \otimes \theta^b) = \theta^b \otimes \theta^a.     \eqno(4.1)
$$
The metric on the fuzzy bundle can be defined by the condition
$$
g(\theta^a \otimes \theta^b) = g^{ab}                              \eqno(4.2)
$$
with $g^{ab}$ the components of the euclidean metric. This defines a
metric on the surface in the commutative limit.

The choice of metric is strongly restricted by the choice of
differential calculus. If we introduce a new basis $\bar\theta^a$
of $\Omega^1(M_n)$ defined by 
$\bar\theta^a = (\delta^a_b + \Lambda^a_b) \theta^b$ with the
$\Lambda^a_b$ arbitrary elements of $M_n$ then of course the
coefficients of the metric change accordingly just as in the commutative
case. The difference lies in the fact that we cannot introduce a new
metric $\bar g$ given by 
$\bar g(\bar\theta^a \otimes \bar\theta^b) = g^{ab}$.  Unless
the $\Lambda^a_b$ lie in the center of the algebra this equation would
be inconsistent since $f g^{ab} = g^{ab} f$ but in general
$$
f \bar g(\bar\theta^a \otimes \bar\theta^b) =
\bar g(f \bar\theta^a \otimes \bar\theta^b) \neq
\bar g(\bar\theta^a \otimes \bar\theta^b f) =
\bar g(\bar\theta^a \otimes \bar\theta^b) f.
$$
A metric on the new calculus can be defined by the condition
$$
g(\theta^{\prime a} \otimes \theta^{\prime b}) = g^{ab}.           \eqno(4.3)
$$
Since they lie in different differential calculi we must consider
$\theta^{\prime a} \neq \bar\theta^a$ if $\Lambda^a_b$ is given by (3.16).

The perturbed linear connection associated to the perturbed differential
calculus is defined by the covariant derivative
$$
D\theta^{\prime a} = - \omega^{\prime a}{}_{bc} 
\theta^{\prime b} \otimes \theta^{\prime c}                        \eqno(4.4)
$$
and a generalized perturbation
$$
\sigma^\prime(\theta^{\prime a} \otimes \theta^{\prime b}) = 
S^{\prime ab}{}_{cd} \theta^{\prime c} \otimes \theta^{\prime d}.
$$
We write as above 
$$
S^{\prime ab}{}_{cd} = \delta^b_c \delta^a_d + S^{ab}_{(1)cd}
$$
and expand also the coefficients of the maps $\tau^\prime$ and 
$\chi^\prime$ introduced in Madore \& Mourad (1996):
$$
T^{\prime ab}{}_{cd} = 2 \delta^a_c \delta^b_d + T^{ab}_{(1)cd}, \qquad
\chi^{\prime a}{}_{bc} = \chi^a_{(1)bc}.
$$
Then the most general expression for $\omega^{\prime a}{}_{bc}$ is 
given by
$$
\omega^{\prime a}{}_{bc} = {1\over 2} C^a{}_{bc} + \chi^a_{(1)bc}
- i \kbar^{-1} x_d S^{ad}_{(1)bc}                                  \eqno(4.5)
$$
and the most general expression for $S^{ab}_{(1)cd}$ is given by
$$
S^{ab}_{(1)cd} = {1\over 2} T^{ab}_{(1)}{}_{(cd)} - 2 P^{ab}_{(1)cd}.
                                                                   \eqno(4.6)
$$
Comparing (4.5) with (3.9) we see that the condition 
$$
\omega^{\prime a}{}_{[bc]} = C^{\prime a}{}_{bc}
$$
that the connection be torsion-free is indeed respected.  The most
`natural' generalized permutation is given by $\tau = 2$.  We have then
$$
\sigma(\theta^{\prime a} \otimes \theta^{\prime b}) = 
\theta^{\prime b} \otimes \theta^{\prime a} - 
2 P^{ab}{}_{(1)cd} \theta^{\prime c} \otimes \theta^{\prime d}.     \eqno(4.7)
$$

A straightforward generalization of the definition of metric
compatibility (Dubois-Violette {\it et al.} 1995, Madore \& Mourad 1996)
can be written in the form
$$
\omega^{\prime a}{}_{bc} + 
\omega^\prime_{cd}{}^e S^{\prime ad}{}_{be} = 0.                    \eqno(4.8)
$$
A short calculations yields that the left-hand side of this equation
is given by
$$
\omega^{\prime a}{}_{bc} + 
\omega^\prime_{cd}{}^e S^{\prime ad}{}_{be} = 
- i \kbar^{-1} x_d (S^{ad}_{(1)bc} + S_{(1)c}{}^d{}_b{}^a) +
(\chi^a_{(1)bc} + \chi_{(1)cb}{}^a +
{1\over 2}  C^e{}_{cd} S^{ad}_{(1)be}).                             \eqno(4.9)
$$
Provided the constraint
$$
C^e{}_{cd} S^{ad}_{(1)be} = C^{ea}{}_d S_{(1)c}{}^d{}_{be}         \eqno(4.10)
$$
is satisfied the condition that the last term of (4.9) vanish yields
the value
$$
\chi^a_{(1)bc} = - {1\over 4} (C^e{}_{cd} S^{ad}_{(1)be} +
C^{ea}{}_d S_{(1)b}{}^d{}_{ce} - C^e{}_{bd} S_{(1)c}{}^{da}{}_e)   \eqno(4.11)
$$
for the coefficients of $\chi$. The condition of metric compatibility 
becomes then
$$
S^{ad}_{(1)bc} + S_{(1)c}{}^d{}_b{}^a = 0.                         \eqno(4.12)
$$
The Equations~(4.10) and (4.12) are to be considered as equations for
the coefficients of $\tau$. There is a solution to these two equations,
given by
$$
T^{ad}_{(1)(bc)} = - 4 P_{(1)(b}{}^{da}{}_{c)}                     \eqno(4.13)
$$
provided that $f_{abc}$ is completely symmetric in all indices. The
expression (4.11) for the coefficients of $\chi$ simplifies to
$$
\chi^a_{(1)bc} = 2 f^a{}_{bc}.                                     \eqno(4.14)
$$
The corresponding coefficients of $\sigma$ are given by
$$
S^{ad}_{(1)bc} = - 2 (P^{ad}_{(1)bc} - 
P_{(1)b}{}^d{}_c{}^a + P_{(1)c}{}^{da}{}_b).                       \eqno(4.15)
$$
There are perhaps other solutions. The final values for the coefficients
of the connection are found by using (4.14) and (4.15) in (4.5). The
resulting expression is formally the same in the commutative limit with
the matrices $x^a$ replaced by coordinates $x^a$.

Using (4.14) and (4.15) we see that the coefficients (4.5) of the 
perturbed linear connection can be written in the form
$$
\omega^{\prime a}{}_{bc} = {1\over 2} C^a{}_{bc} + 2 f^a{}_{bc}
+ 2 x^\prime_d (f_b{}^{de} C^a{}_{ce} + f_b{}^{ae} C^d{}_{ce} -
f_{bc}{}^e C^{da}{}_e).                                            \eqno(4.16)
$$

We have constructed a linear connection on a bundle over the surface. It
will yield upon reduction {\it \`a la} Kaluza-Klein a linear connection,
an electromagnetic potential and a scalar field on the surface itself.
From (2.16) we see that the electromagnetic field will have unit
magnetic charge. We see here also that the electromagnetic potential can
be identified with the component of the frame normal to the surface.

\section{Discussion}

One can formally write a curvature associated to $(D^\prime,
\sigma^\prime)$ but the definition of curvature is not satisfactory. The
main interest of curvature in the case of a smooth manifold $V$ is the
fact that it is local. Riemann curvature can be defined as a map
$$
\Omega^1({\cal C}(V)) {\buildrel R \over \longrightarrow} 
\Omega^2({\cal C}(V)) \otimes_{{\cal C}(V)} \Omega^1({\cal C}(V)).
$$
If $\xi \in \Omega^1({\cal C}(V))$ then $R(\xi)$ is local; at a given
point it depends only on the value of $\xi$ at that point. This can be
expressed as a bilinearity condition; the above map is a 
${\cal C}(V)$-bimodule map.  If $f \in {\cal C}(V)$ then
$$
f R(\xi) = R(f\xi), \qquad  R(\xi f) = R(\xi) f.
$$
In the noncommutative case bilinearity is the natural (and only
possible) expression of locality. It has not yet been possible to
enforce it in a satisfactory manner. We refer to Dubois-Violette {\it et
al.} (1996) for a recent discussion.

If we compare the expression (3.11) for the induced Poisson structure
with the expression (4.16) for the linear connection we see that there is
definitely a connection between the two. Further evidence of this is
given in another article (Madore 1997).  The exact extent to which
one is determined by the other is not clear.

Although we are primarily interested in the matrix version of surfaces
as an model of an eventual noncommutative theory of gravity they have a
certain interest in other, closely related, domains of physics. Without
the differential calculus the fuzzy sphere is basically just an
approximation to a classical spin $r$ by a quantum spin $r$ given by
(2.15) with $\hbar$ in lieu of $\kbar$.  It has been extended in various
directions under various names and for various reasons by Berezin
(1975), de~Wit {\it et al.} (1988), Hoppe (1989), Fairlie {\it et al.}
(1989), Floratos {\it et al.} (1989), Cahen {\it et al.} (1990) and
Bordemann {\it et al.} (1991).  In order to explain the finite entropy
of a black hole it has been conjectured, for example by 't~Hooft (1996),
that the horizon has a structure of a fuzzy 2-sphere since the latter
has a finite number of `points' and yet has an $SO_3$-invariant
geometry.  Matrix descriptions of general $p$-dimensional hypersurfaces
were first proposed in connection with Dirichlet $p$-branes and their
relation to $M$-theory (Banks {\it et al.} 1996). Attempts to endow
them with a smooth differential structure has been made (Madore
1996, Grosse {\it et al.} 1997a).

\section*{Acknowledgments} The author would like to thank J. Gratus and 
L.A. Saeger for interesting conversations.

\parindent=0cm
\tolerance=1000
\parskip 4pt plus 1pt

\section*{References}

Banks T., Fishler W., Shenker S.H., Susskind L. 1996, 
{\it $M$ Theory as a Matrix Model: a Conjecture}, hep-th/9610043.

Bordemann M., Hoppe J., Schaller P., Schlichenmaier M. 1991,
{\it $gl(\infty )$ and Geometric Quantization}, 
Commun. Math. Phys. {\bf 138} 209.

Berezin F. 1975, {\it General Concept of Quantization},
Commun. Math. Phys. {\bf 40} 153.

Cahen M., Gutt S., Rawnsley J. 1990, 
{\it Quantization of K\"ahler Manifolds I}, 
Jour. of Geom. and Phys. {\bf 7} 45.

Carow-Watamura U., Watamura S. 1997, {\it Chirality and Dirac Operator
on Noncommutative Sphere}, Commun. Math. Phys. {\bf 183} 365.

Connes A.  1994, {\it Noncommutative Geometry}, Academic Press.

de Wit B., Hoppe J., Nicolai H. 1988, 
{\it On the Quantum Mechanics of Supermembranes}, 
Nucl. Phys. {\bf B305} 545.

Dimakis A., Madore J. 1996, {\it Differential Calculi and Linear
Connections}, J. Math. Phys. {\bf 37} 4647.

Dubois-Violette M. 1988, 
{\it D\'erivations et calcul diff\'erentiel non-com\-mutatif}, 
C. R. Acad. Sci. Paris {\bf 307} S\'erie I 403.

Dubois-Violette M., Kerner R., Madore J. 1989, 
{\it Gauge bosons in a noncommutative geometry}, 
Phys. Lett. {\bf B217} 485; 
{\it Classical bosons in a noncommutative geometry}, 
Class. Quant. Grav. {\bf 6} 1709.

Dubois-Violette M., Madore J., Masson T., Mourad J. 1995, 
{\it Linear Connections on the Quantum Plane}, 
Lett. Math. Phys. {\bf 35} 351.

--- 1996, {\it On Curvature in Noncommutative Geometry}, 
J. Math. Phys. {\bf 37} 4089.

Fairlie D.B., Fletcher P., Zachos C.K. 1989, 
{\it Trigonometric Structure Constants for New Infinite Algebras},
Phys. Lett. {\bf B218} 203.

Floratos E.G., Iliopoulos J., Tiktopoulos G. 1989, 
{\it A Note on $SU(\infty)$ Classical Yang-Mills Theories}, 
Phys. Lett. {\bf B217} 285.

Grosse H., Madore J. 1992, {\it A Noncommutative Version of the Schwinger
Model}, Phys. Lett. {\bf B283} 218.

Grosse H., Pre\v snajder P. 1993, {\it The Construction of
Noncommutative Manifolds Using Coherent States}, 
Lett. in Math. Phys. {\bf 28} 239.

Grosse H. Klim\v c\'\i k C. Pre\v snajder P. 1997a {\it On 4D Field Theory
in Non-Commutative Geometry}, Commun. Math. Phys. {\bf 180} 429.

Grosse H., Klim\v c\'\i k C., Pre\v snajder P. 1997b, {\it Field Theory on a
Supersymmetric Lattice}, Commun. Math. Phys. {\bf 185} 155.

Hoppe J. 1989, {\it Diffeomorphism groups, Quantization and $SU(\infty)$},
Int. J. Mod. Phys. {\bf A4} 5235.

Madore J. 1992, {\it The Fuzzy Sphere}, Class. Quant. Grav. {\bf 9} 69.

--- 1995, {\it An Introduction to Noncommutative Differential Geometry
and its Physical Applications}, Cambridge University Press.

Madore J. 1996, {\it Linear Connections on Fuzzy Manifolds}, 
Class. Quant. Grav. {\bf 13} 2109.

Madore J. 1997, {\it On Poisson Structure and Curvature}
Preprint LPTHE Orsay 97/25, gr-qc/9705083.

Madore J., Mourad J. 1996, 
{\it Quantum Space-time and Classical Gravity}, 
Preprint, LPTHE Orsay 96/56, gr-qc/9607060.

Madore J., Masson T., Mourad J. 1995, 
{\it Linear Connections on Matrix Geometries}, 
Class. Quant. Grav. {\bf 12} 1429.

Mourad. J. 1995, {\it Linear Connections in Non-Commutative Geometry},
Class. Quant. Grav. {\bf 12} 965.

't Hooft G. 1996 {\it Quantization of point particles in
(2+1)-dimensional gravity and spacetime discreteness},
Class. Quant. Grav. {\bf 13} 1023.

\end{document}